\documentclass[mathpazo]{cicp}

\usepackage[dvips]{graphicx}
\usepackage{amsfonts,amssymb,amsbsy,amsmath,amsthm}
\usepackage[english]{babel}
\usepackage[utf8]{inputenc}
\usepackage{cite}
\usepackage{xcolor}
\usepackage{hyperref}

\begin{document}

\title{Lattice Boltzmann Simulations of Two Linear Microswimmers Using the Immersed Boundary Method}

\author[]{D. Geyer\affil{1}, S. Ziegler\affil{2}, A. Sukhov\affil{1}, M. Hubert\affil{2}, A.-S. Smith\affil{2}, O. Aouane\affil{1}, P. Malgaretti\affil{1} and J. Harting\affil{1,3}\corrauth}
\address{\affilnum{1} Helmholtz Institute Erlangen-N\"{u}rnberg for Renewable Energy (IEK-11), Forschungszentrum J\"{u}lich, Cauerstra{\ss}e 1, 91058 Erlangen, Germany \\
\affilnum{2}PULS Group, Department of Physics, Interdisciplinary Center for Nanostructured Films, Friedrich-Alexander-Universit\"at  Erlangen-N\"{u}rnberg, Cauerstra{\ss}e 3, 91058 Erlangen, Germany\\
\affilnum{3} Department of Chemical and Biological Engineering and Department of Physics, Friedrich-Alexander-Universit\"at Erlangen-N\"urnberg, F\"{u}rther Stra{\ss}e 248, 90429 N\"{u}rnberg, Germany}
\email{{\tt j.harting@fz-juelich.de} (J.~Harting)}

\begin{abstract}
The performance of a single or the collection of microswimmers strongly depends on the hydrodynamic coupling among their constituents and themselves.
We present a numerical study for a single and a pair of microswimmers based on lattice Boltzmann method (LBM) simulations.   
Our numerical algorithm consists of two separable parts. 
Lagrange polynomials provide a discretization of the microswimmers and the lattice Boltzmann method captures the dynamics of the surrounding fluid. 
The two components couple via an 
immersed boundary method. 
We present data for a single swimmer system and our data also show the onset of collective effects and, in particular, an overall velocity increment of clusters of swimmers.
\end{abstract}


\maketitle

\section{Introduction}
\label{introduction}
In his seminal work, Purcell pointed out that the properties of microscopic objects placed in fluids are significantly different from their macroscopic counterparts~\cite{Purcell1977}. 
In particular, Purcell showed that in order to swim in the low Reynolds number regime, a micrometric swimmer has to move its parts in such a manner as to break the time inversion symmetry.  
This fact led to the well-known ``scallop theorem'',  which states that in order to attain self-propulsion in the low Reynolds number regime, at least two degrees of freedom are needed.
Since then, numerous attempts have been done to elucidate the dynamics of microswimmers by means of theoretical models \cite{Blake1971, Felderhof2006, EarlPooleyRyder2007, Najafi2004, Golestanian2008, AvronKennethOaknin2005, DowntonStark2009,  Pande2015, Ziegler2019, Wang2019, DaddiLisickiHoell2018, DaddiKurzthalerHoell2019, RizviFarutinMisbah2018}, experimental setups \cite{DreyfusBaudryRoper2005, Ahmed2015, Grosjean2015, Grosjean2016,Palagi2016, GrosjeanHubertCollard2018, ZhengDaiWang2017, HamiltonPetrovWinlove2017, BryanShelleyParish2017, HamiltonGilbertPetrov2018, Collard2020} and numerical  simulations \cite{Gompper2020, BeckerKoehlerStone2003, ZottlStark2012, PicklGoetzIglberger2012, PicklPandeKoestler2017, Sukhov2019, Peter2020}.
In particular, in Ref.~\cite{Najafi2004} Najafi and Golestanian proposed a theoretical model that precisely fulfills the requirement of Purcell. 
Indeed, they offered a very simple swimmer composed of three aligned solid spherical particles suspended in a viscous fluid and actuated internally by changing the distances between neighboring particles. 
In a Newtonian fluid (see Refs.~\cite{Hubert2021, Qiu2014} for a theoretical and experimental extension of the problem in the case of the underdamped regime and non-Newtonian fluids), the dynamics of such a swimmer is fully determined by the two degrees of freedom of the swimmer, namely, the two distances among subsequent beads \cite{Golestanian2008}. 
Later, a variation of this swimmer has been proposed, where harmonic springs connect neighboring particles and external forces drive the swimmer \cite{Felderhof2006, Pande2015}.
Once actuated with a proper protocol, the motion of the three beads of the swimmer results to be non-reciprocal and hence leads to a net displacement.\\
Recently, the focus of theoretical research has shifted towards swimming in complex environments like channels~\cite{Elgeti2015, Malgaretti2017} or near fluid interfaces~\cite{Sukhov2019, Malgaretti2021}, and the question of collective swimmer dynamics has become of major interest. 
The present work aims at understanding the collective dynamics of several microswimmers~\cite{Buzhardt2019, Ziegler2021}.
For this purpose, we consider a relatively simple situation of two linear microswimmers in different configurations. 
Additionally, we exploit the suitability of the employed simulation method for this task. 
The structure of the manuscript is as follows:
In Sec.~\ref{s: problem}, we present the theoretical model and the numerical implementation.
In Sec.~\ref{s: results}, we present our numerical results for a single and two microswimmers in diverse arrangements, which are then discussed in Sec.~\ref{s: discussion}. 
Finally, in Sec.~\ref{s: conclusions} we provide some concluding remarks.

\section{Model}
\label{s: problem}
\begin{figure}[ht!]
  \centering
  \includegraphics[width=0.8\linewidth]{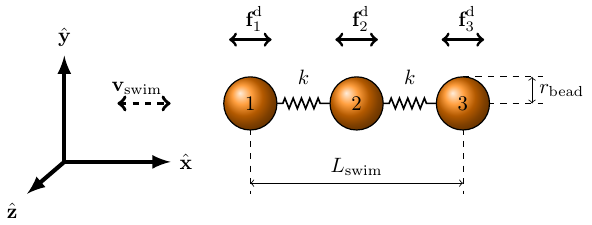}
  \caption{Scheme of the linear, three bead swimmer connected with two springs.
  The motion direction $v_\text{swim}$ depends on the interplay of the amplitudes and phases of external forces as detailed in \cite{Grosjean2015}. 
  Distances between beads are measured from the respective centers of mass.}
 \label{fig: bead-spring model}
\end{figure}

We focus on investigating the behavior of a single and a pair of bead-spring microswimmers in a resting Newtonian fluid. 
Each microswimmer consists of three aligned equal beads connected with springs which is a modification~\cite{Pande2015} of the model proposed by Najafi and Golestanian~\cite{Najafi2004}.
The distances between the beads represent the two degrees of freedom necessary to attain self-propulsion, as per the scallop theorem~\cite{Purcell1977} and depicted in Fig.~\ref{fig: bead-spring model}. 
The microswimmer particles consist of rigid spherical shells filled with a Newtonian fluid with the same viscosity and density as the external fluid. 
The fluid-particle interactions are incorporated into the model through the immersed boundary method (IBM)~\cite{Peskin2002}. 
The flow field in the entire computational domain is computed using the lattice Boltzmann method (LBM).

\subsection{Fluid dynamics}

In this study, we use the single relaxation time LBM with a Bhatnagar, Gross, Krook \cite{Bhatnagar1954} collision operator to solve for the flow field on an Eulerian frame in the weakly compressible limit. 
The macroscopic fluid dynamics are recovered from the mesoscopic Boltzmann equation as detailed in Ref.~\cite{benzi1992lattice, succi2001}. The LBM consists of two steps, namely collision and advection. We adopt a three-dimensional 19 velocities lattice (D3Q19) model and represent lengths in units of the lattice spacing $\Delta x$ and times in units of the integration time step $\Delta t$. The time evolution of the distribution functions $f_i(\textbf{x},t)$ is obtained by solving the discrete lattice Boltzmann equation in velocity space such that
\begin{equation}
  f_i({\mathbf{x}}+\mathbf{c}_i\Delta t, t + \Delta t) - f_i({\mathbf{x}},t) = -\frac{\Delta t}{\tau} [f_i({\mathbf{x}},t)-f_i^{eq}({\mathbf{x}},t)],
  \label{eq:lbe}
\end{equation}
where $f_i$ describes the discrete probability of finding a fluid particle at position $\mathbf{x}$ and time $t$ moving with velocity $\mathbf{c}_i$ for $i = 1, \ldots, 19$. $\tau$ is the relaxation time or rate at which the system relaxes toward a local equilibrium distribution function $f_i^{eq}$ corresponding to the truncated expansion of the Maxwell-Boltzmann distribution for the velocities in an ideal gas. $f_i^{eq}$ is expressed as
\begin{equation}
  f_i^{eq} = \omega_i \rho \bigg[ 1 + \frac{\mathbf{c}_i \cdot \mathbf{u}^{eq}}{c_s^2} - \frac{ \left( \mathbf{u}^{eq} \cdot \mathbf{u}^{eq} \right) }{2 c_s^2} + 
  \frac{ \left( \mathbf{c}_i \cdot \mathbf{u}^{eq} \right)^2}{2 c_s^4}  \bigg],
    \label{eq:lbm:equilibrium}
\end{equation}
where $\rho$ and $\mathbf{u}^{eq}$ are the macroscopic number density and the equilibrium velocity, and $c_s=1/\sqrt{3}\Delta x / \Delta t$ is the lattice speed of sound. $\omega_i$ are the lattice weights which read as $1/3$, $1/18$ and $1/36$ for $i=1$, $i=2\dots7$, and $i=8\dots19$, respectively. $\rho$ and $\mathbf{u}^{eq}$ are obtained from the moments of the particle distribution function $f_i$ such that
\begin{equation}
   \rho = \sum_{i=1}^{19} f_i ({\mathbf{x}},t), \quad \text{and} \hspace{0.25cm}  \rho\mathbf{u}^{eq} = \sum_{i=1}^{19} f_i (\mathbf{x},t)\mathbf{c}_i + \tau \mathbf{F},
   \label{eq:sc}
\end{equation}
where $\mathbf{F}$ is the total force applied by the particles on the fluid.
Note that since we are using the Shan and Chen scheme \cite{shan1993lattice} to include external forces on the fluid, the equilibrium velocity $(\mathbf{u}^{eq})$ used in Eq. \eqref{eq:lbm:equilibrium} is different from the physical velocity of the fluid nodes $(\mathbf{u})$ which is defined as
\begin{equation}
    \rho \mathbf{u} = \sum_{i=1}^{19} f_i ({\mathbf{x}},t)\mathbf{c}_i + \frac{\Delta t}{2}\mathbf{F}.
\end{equation}
Finally, the dynamic viscosity $\mu = \nu/\rho_0$ with $\nu$ being the kinematic viscosity, reads as
\begin{equation}
    \mu = \rho_0 c_s^2\left(\tau - \frac{\Delta t}{2}\right),  
\end{equation}
where $\rho_0 = m \rho$ is the mass density with $m$ here being the molecular mass of the fluid particles. For convenience, $\Delta x$, $\Delta t$, and $\rho_0$ are set to unity in this study.

\subsection{Microswimmer model}
\label{ss: microswimmer model}

We consider a force-driven microswimmer consisting of three beads of equal size with two harmonic springs as depicted in Fig.~\ref{fig: bead-spring model}.
This is one of the simplest possible microswimmer models~\cite{Felderhof2006}. A driving force $\textbf{f}^{d}$ exerted on each bead of the microswimmer is prescribed as \cite{Pande2015}
\begin{align}
    \textbf{f}_\text{1}^{d}(t) &= A\ \sin(\omega t)\ \hat{\mathbf{x}}, \nonumber \\
    \textbf{f}_\text{2}^{d}(t) &= - \mathbf{f}_\text{1}^{d}(t) - \mathbf{f}_\text{3}^{d}(t)\text{ and }\\
    \textbf{f}_\text{3}^{d}(t) &= B\ \sin(\omega t + \alpha)\ \hat{\mathbf{x}} \text{ with } \alpha \in [-\pi, \pi]. \nonumber
    \label{eq: ext forces}
\end{align}
Here $A$ and $B$ are non-negative amplitudes of the time-dependent driving forces $\mathbf{f}_1^{d}(t)$ and $\mathbf{f}_3^{d}(t)$ applied along the $x$-axis to the outer beads at the frequency $\omega = 2 \pi / T$. A crucial condition for self-propelled objects is that the sum over all external driving forces equals zero at all times.
For simplicity, we assume that the two harmonic springs connecting the beads are identical with a stiffness $k$ and an equilibrium length $L_\text{spring}$ which implies that the total length of a single swimmer is $\sim 2 L_\text{spring}$.
The spring forces on the beads are given by 
\begin{align}
    \textbf{f}_\text{1}^{s}(t) &= -k (|\mathbf{R}_1(t) - \mathbf{R}_2(t)| - L_\text{spring}) \frac{\mathbf{R}_1(t) - \mathbf{R}_2(t)}{|\mathbf{R}_1(t) - \mathbf{R}_2(t)|}, \nonumber \\
    \textbf{f}_\text{2}^{s}(t) &= - \mathbf{f}_\text{1}^{s}(t) - \mathbf{f}_\text{3}^{s}(t),\\
    \textbf{f}_\text{3}^{s}(t) &= -k (|\mathbf{R}_3(t) - \mathbf{R}_2(t)| - L_\text{spring}) \frac{\mathbf{R}_3(t) - \mathbf{R}_2(t)}{|\mathbf{R}_3(t) - \mathbf{R}_2(t)|}, \nonumber 
\end{align}
with $\mathbf{R}_i$ the position vector of bead $i$.\\
Our model is limited to small forces such that the assumptions for the Stokes regime remain valid.
For this purpose, it is critical to differentiate between two Reynolds numbers
\begin{align}
    Re_\text{swim} = \frac{{\rho_0} \cdot |\textbf{v}_\text{swim}| \cdot 2\ L_\text{spring}}{\mu} \quad \text{ and } \quad
    Re_\text{bead} = \frac{{\rho_0} \cdot |\textbf{v}_\text{max fastest bead}| \cdot 2\ r_\text{bead}}{\mu},
\end{align}

and ensure that both $Re_\text{swim}$ and $Re_\text{bead}$ are small enough.\\
Each spherical bead in our microswimmer is generated from an icosahedron that is refined recursively until obtaining a sufficiently smooth surface described by a triangular mesh \cite{Aouane2018}.  
Our fluid-filled particles are modeled using a strain-hardening constitutive law known as the Skalak strain energy \cite{Skalak1973,Skalak1973b}, which is written as 
\begin{equation}
  \mathcal{W}_{sk} = \frac{\kappa_s}{4}\oint[I_1^2 + 2 I_1 -2 I_2 + C I_2^2]dA,
\end{equation}
where the eigenvalues $\lambda_1$, $\lambda_2$ of the displacement tensor define deformation invariants $I_1=\lambda_1^2 + \lambda_2^2 -2$ and $I_2 = \lambda_1^2 \lambda_2^2 -1$, and $C$ is a constant parameter controlling the extensibility of the membrane. The area dilatation modulus $\kappa_a$ is defined such as $\kappa_a/\kappa_s= 1 + 2C$, with $\kappa_s$ being the shear elastic modulus. The deformations are evaluated using a linear finite element method \cite{Kruger2011}. In addition to resistances to shear elasticity and area dilatation, our particle membrane can withstand out of plane deformations (i.e. bending). The curvature energy is accounted for via the Helfrich free energy
\begin{equation}
   \mathcal{W}_b = \frac{\kappa_b}{2}\oint [2H]^2 dA + {\kappa_g}\oint K \, dA,
\end{equation}
where $H = \frac{1}{2}\sum_{k=1}^2 \bar{C}_k $, and $K=\prod_{k=1}^2 \bar{C}_i$ are the mean and Gaussian curvatures. $\bar{C}_1$ and $\bar{C}_2$ are the two principal curvatures. $\kappa_b$ and $\kappa_g$ are the bending and Gaussian curvatures moduli. The discretization of the bending energy follows the approach of Kantor and Nelson \cite{Kantor1987} for flat triangulated meshes.
The volume conservation of the capsule is enforced using a penalty function reading as
\begin{equation}
\mathcal{W}_v = \frac{\kappa_v}{2}\frac{[V-V_0]^2}{V_0},
 \label{eq:vol_cons}
\end{equation}
where $V_0$ is the reference volume of the stress-free particle, and $\kappa_v$ is a constant parameter. 
$\mathbf{r}_{i,j}$ is the position of the $j$-th mesh node belonging to the $i$-th particle.
The forces resulting from bending, shear elasticity and the constraint on the volume are evaluated using the principle of virtual work such that $\mathbf{f}^{part}_{i,j}(\mathbf{r}_{i,j}) =-\frac{\partial \mathcal{W}}{\partial \mathbf{r}_{i,j}}$ with $\mathcal{W} = \mathcal{W}_{sk} + \mathcal{W}_b + \mathcal{W}_v$. 
The driving and spring forces are distributed over the mesh nodes $j$ of the corresponding particle such that $\mathbf{f}_{i,j}^{\alpha}(\mathbf{r}_{i,j}) \equiv  \mathbf{f}_i^{\alpha}(\mathbf{r}_{i,j}) / N_v$, with $N_v$ being the total number of mesh nodes on particle $i$, and the index $i$ runs over the number of particles, here from $1$ to $3$. The superscript $\alpha$ stands for either {\it driving} $d$ or {\it spring} $s$. 
The total nodal force ($\mathbf{f}_{i,j}^{tot}$) on particle $i$ mesh node $j$, which includes the contribution of the external forces, is defined as
\begin{equation}
    \mathbf{f}_{i,j}^{tot}(\mathbf{r}_{i,j}) = \mathbf{f}_{i,j}^{part}(\mathbf{r}_{i,j}) + \mathbf{f}_{i,j}^{d}(\mathbf{r}_{i,j}) + \mathbf{f}_{i,j}^{s}(\mathbf{r}_{i,j}).
\end{equation}
By choosing the appropriate values of $\kappa_s$, $C$, $\kappa_b$ and $\kappa_v$, it is possible to work in the small deformation regime where the particles are quasi-rigid. The chosen values are given in Sec.~\ref{s: results}.
\subsection{Fluid-Particle interaction}
\label{ss: fluid-particle interaction}
The immersed boundary method (IBM) is a fluid-structure coupling method that was first introduced by Peskin in the early seventies of the last century to model the flow patterns around heart valves~\cite{Peskin1973}. 
The IBM involves both Eulerian and Lagrangian quantities.
The Eulerian variables exist on a Cartesian grid representing the fluid region while the Lagrangian variables are based on a moving curvilinear mesh representing the interface. 
A smoothed approximation of the Dirac delta function is used to transfer data from one mesh to the other. 
The distribution of the particle nodal forces to the neighboring fluid nodes reads as
\begin{equation}
  \mathbf{F}(\mathbf{x},t) = \sum_j\sum_i \mathbf{f}_{i,j}^\textnormal{tot} \Delta(\mathbf{x}-\mathbf{r}_{i,j}),
  \label{eq:spreading}
\end{equation}
where $\sum_j$ is a sum over all the membrane nodes $\mathbf{r}_{i,j}$ located within an interpolation range from the fluid node $\mathbf{x}$. 
At this point, we reason in terms of interactions between Lagrangian mesh nodes (not in terms of individual particles) and Eulerian fluid nodes. $\Delta$ is the smoothed discrete Dirac delta function, and $\mathbf{F}(\mathbf{x},t)$ is the force density acting on the fluid at the Eulerian node $\mathbf{x}(x_1,x_2,x_3)$ due to the contributions of the membrane total force ($\mathbf{f}^\textnormal{part}$), the external driving force ($\hat{\mathbf{f}}^\textnormal{d}$) and the spring force ($\hat{\mathbf{f}}^\textnormal{s}$).
Similarly, the interpolation of the velocity of the neighboring fluid nodes onto a membrane Lagrangian node $\mathbf{r}(r_1,r_2,r_3)$ is performed as 
\begin{equation}
  \dot{\mathbf{r}}_{i,j} = \sum_x \mathbf{u}(\mathbf{x}) \Delta(\mathbf{x}-\mathbf{r}_{i,j}),
  \label{eq:interpolation}
\end{equation}
where $\sum_x$ is a sum over all the fluid nodes within an interpolation range from the membrane node $\mathbf{r}_{i,j}$. $\Delta(\mathbf{x}-\mathbf{r}_{i,j})$ is then replaced by a two-point linear interpolation function as detailed in \cite{Kruger2012,Aouane2021}. 
Eqs. \eqref{eq:spreading} and \eqref{eq:interpolation} describe the spreading of the interfacial forces to the surrounding fluid nodes and the interpolation of the fluid velocity to the deformable interface. The particle forces are included directly in the fluid node velocity using the method proposed by Shan and Chen \cite{shan1993lattice} as described in Eq. \eqref{eq:sc}.\\
The IBM belongs to the class of the front-tracking methods where the sharp interface is explicitly known through a set of Lagrangian marker points acting on the fluid via body forces and moving with the same velocity as the ambient fluid, thus enforcing the no-slip boundary condition. 
The deformation of the interface is governed by the chosen strain-stress constitutive law and not by the IBM itself, making it a very popular method to simulate biological fluid dynamics (e.g., blood flow) \cite{Zhang2007, Crowl2010, kaoui2011two, thiebaud2014prediction, kruger2014deformability}.

\subsection{Analytical calculation}
To compare the results of the lattice Boltzmann method, we make use of an analytical framework~\cite{Ziegler2021} where the hydrodynamic interactions between the particles are calculated using either the Oseen or Rotne-Prager approximation~\cite{Dhont1996}. The equation of motion of the bead-spring system reads
\begin{equation}
    \dot{\mathbf{R}}_i (t) = \frac{1}{6 \pi \mu r_\text{bead}} (\mathbf{f}^d_i (t) + \mathbf{f}^s_i (t)) + \sum_{j \neq i} \hat{T} ( \mathbf{R}_i(t) - \mathbf{R}_j(t)) \cdot (\mathbf{f}^d_j(t) + \mathbf{f}^s_j (t)),
    \label{eq:EOM}
\end{equation}
with particle indices $i, j$, the position $\mathbf{R}_i$ of particle $i$, the bead radius $r_\text{bead}$ and the spring force $\mathbf{f}^s_i$ on bead $i$.
The tensor $\hat{T}$ is given by either the Oseen tensor, 
\begin{equation}
    \hat{T} (\mathbf{r}) = \frac{1}{8 \pi \mu |\mathbf{r}|} \left(\hat{1} + \frac{\mathbf{r} \otimes \mathbf{r}}{\mathbf{r}^2} \right), 
\end{equation}
or the Rotne-Prager tensor \cite{PicklPandeKoestler2017}
\begin{equation}
    \hat{T} (\mathbf{r}) = \frac{1}{8 \pi \mu |\mathbf{r}|} \left(\hat{1} + \frac{\mathbf{r} \otimes \mathbf{r}}{\mathbf{r}^2} \right) + \frac{r_\text{bead}^2}{12 \pi \mu |\mathbf{r}|^3} \left(\hat{1} - 3\frac{\mathbf{r} \otimes \mathbf{r}}{\mathbf{r}^2} \right). 
    \label{eq: Rotne-Prager}
\end{equation}
Here, $\hat{1}$ denotes the $3 \times 3$ unit matrix and $\otimes$ the tensor product. 
The Oseen tensor assumes point forces, and the Rotne-Prager tensor correctly describes the flow field around a spherical particle.
Eq.~\eqref{eq: Rotne-Prager} also accounts for the correction to the interaction term arising as a consequence of Faxen's law \cite{KimKarrila1991}. 
As a result, the prefactor of the second term in Eq.~\eqref{eq: Rotne-Prager} is $r^2_\text{bead}/(12 \pi \mu |\bf{r}|^3)$ \cite{PicklPandeKoestler2017}.  
The equation of motion \eqref{eq:EOM} is then solved numerically, and the swimmer velocities extracted from the particle trajectories are readily compared to those obtained in the LBM simulations.

\section{Results}
\label{s: results}
We consider a fully periodic simulation domain of size $1024 \times 256 \times 256$.
Each microswimmer is formed by three beads of radius $r_\text{bead} = 5$ and connected with two springs with an equilibrium length of $L_\text{spring} = 36$ and a spring constant $k=2\cdot 10^{-2}$. 
The amplitudes of the time-dependent driving forces $\mathbf{f}_1^d(t)$ and $\mathbf{f}_3^d(t)$ are chosen as $A=10^{-1}$ and $B=5 \cdot 10^{-3}$, respectively. 
The period and the phase shift are set to $T=2\cdot 10^{4}$ and $\alpha=\pi/2$. 
Since the size of the beads is quite smaller than the total length of the microswimmer, then, as mentioned earlier, we can approximate the total length of the swimmer by the sum of the lengths of the two springs at equilibrium $\sim 2 L_\text{spring}$.
We choose $\kappa_s=5\cdot 10^{-2}$, $\kappa_a=1$, $\kappa_b=1.25 \cdot 10^{-3}$ and $\kappa_v=1$ such that the particles remain spherical during the swimmer motion. 
The initial fluid density $\rho_0 = 1$. A typical  simulation run requires approximately 18000 CPU hours.

We calculate the swimmer velocity by the average position of the center of mass (cm) 
\begin{align}
    \textbf{v}_\text{cm} = \frac{\textbf{x}_\text{cm}(t_0 + T) - \textbf{x}_\text{cm}(t_0)}{T},
    \quad \text{ with } \textbf{x}_\text{cm} = \frac{1}{3} \sum_\text{i = 1}^3 \textbf{x}_\text{i}.
    \label{eq:vel_swim}
\end{align}

\subsection{Single microswimmer}

\begin{figure}
    \centering
    \includegraphics[width=1\linewidth]{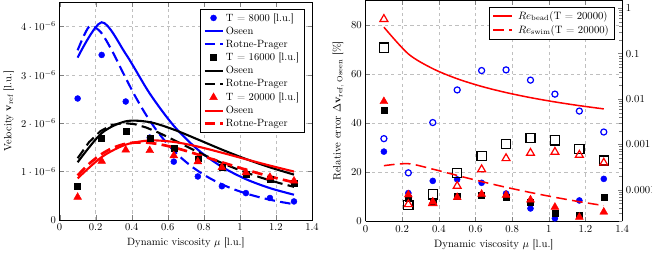}
    \caption{Left: Velocity $\textbf{v}_\text{ref}$ of a single microswimmer (reference swimmer) for different external driving force periods $T$ (see legend). Points stand for LBM data, solid (dashed) lines for analytical predictions based on the Oseen~\cite{Pande2017} (Rotne-Prager~\cite{Ziegler2019}) tensor. 
    Right: relative differences $\Delta \textbf{v}_\text{ref} = |\textbf{v}_\text{ref} - \textbf{v}_\text{theo}|/ \textbf{v}_\text{theo}$ between numerical data and analytical, Oseen  (open points) and Rotne-Prager (solid points) predictions. The red lines report the magnitude of the Reynolds number for the fastest bead (solid) and the swimmer in total (dashed) in the case of $T = 20000$ [l.u.]. The figure shows a better match between LBM data and Rotne-Pragner tensor than LBM data and Oseen tensor. The Reynolds number of the fastest bead plays an essential role by comparing LBM, Rotne-Prager and Oseen approaches. The higher external driving force period implies a decreased Reynolds number, so the tensor approaches fit better to our simulation data.}
    \label{fig: s1}
\end{figure}

\begin{figure}
  \centering
  \includegraphics[width=0.9\linewidth]{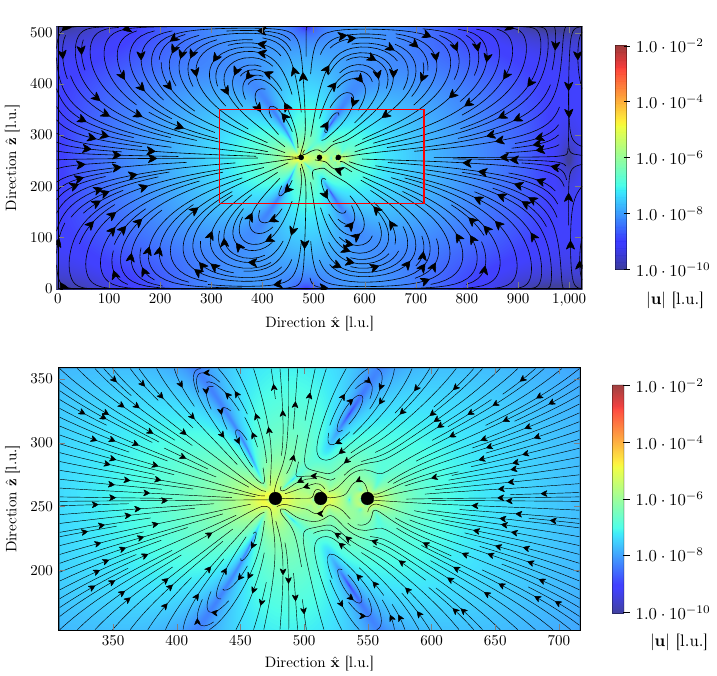}
  \caption{Averaged fluid velocity field for a single microswimmer for dynamic viscosity $\mu = 1/6$ and fluid density $\rho_0 =1$. The image above shows the full box and the image below zooms on the microswimmer.
  The fluid flow (black arrows) surrounding a single microswimmer (black beads) is shown whereby it is averaged over one swimming period in the stationary case. 
  The absolute values of the fluid velocity $\textbf{u}$ (black arrows) are shown by the different colors.}
  \label{fig: s1 vel field}
\end{figure}

First, we characterize the performance of a single microswimmer. 
Long-time simulations over 50 swimming periods $T$ show that the stationary swimming velocity is reached after $\sim 12$ swimming periods. 
Accordingly, the average velocities are defined via Eq.~\eqref{eq:vel_swim} after such a transient period. 
In particular, we focus on the dependence of the swimmer's velocity on the dynamic viscosity of the fluid $\mu$.\\
As expected, we retrieve a non-monotonous velocity dependence of the swimmer on $\mu$, as reported in Ref.~\cite{Pande2017}, for diverse values of the period of the forcing $T$. In particular, the velocity of a single swimmer obtained from the LB simulations compares well to analytical results (Fig. \ref{fig: s1}).
To assess the origin of the quantitative mismatch between our simulation and analytical results using the Oseen tensor, we look at the value of the Reynolds number associated with the swimmer as a whole and with a single bead at its maximum speed. 
Indeed, the Oseen approach of Ref.~\cite{Pande2017} assumes vanishing Reynolds number. 
In contrast, the solid and dotted lines of Fig.~\ref{fig: s1} show that while the Reynolds number of the swimmer is relatively small ($Re_{swimm}\sim 10^{-4}$), the Reynolds number associated with the beads at maximum speed is larger by two orders of magnitude ($Re_{bead} \sim 10^{-2}$). 
Indeed, recent works have shown that when the Reynolds number approaches unity, swimming protocols that exploit inertia can become operational~\cite{Hubert2021, Choudhary2022}.
However, the relative mismatch between the Oseen predictions and the numerical results suggests that fluid inertia may not be the only cause of such a discrepancy. 
An additional assumption of the analytical model is that the fluid velocity adjusts instantaneously to the force, whereas in the LBM simulations for a length $L$, such a time characterized by $\bar{\tau}=L^2/\nu$ is finite and it is represented by the relaxation time $\tau$ in the BGK collision operator~\cite{Bhatnagar1954}. 
On the other hand, also the analytical predictions are derived under approximations. 
In fact, the Oseen approach assumes point particles and it does not properly account for the spherical shape of the particles. 
Accordingly, we compare our data also with a more refined approach based on the Rotne-Prager tensor~\cite{Ziegler2019}. 
Interestingly, Fig.~\ref{fig: s1} shows a better agreement between the LBM simulations and the predictions based on the Rotne-Prager as with the Oseen model~\cite{Pande2017}.\\
Finally, the analytical-numerical mismatch can be due to the periodic boundary conditions that we implement in the LBM simulations compared to the unbound fluid considered in the analytical models. 
The periodic boundary conditions induce spurious interactions of the swimmer with its image. 
Accordingly, to assess this issue, we have checked that the self-interaction across the boundaries is lower than one percent, ruling it out as a possible reason for the mismatch with the analytical results.
This crucial point is also valid for the case of two microswimmers considered below.\\
Fig.~\ref{fig: s1 vel field} shows the velocity field, averaged over a period of the external forces acting on the swimmer. 
The velocity profile of this swimmer resembles that of a force dipole in a far-field expansion. 

\subsection{Two aligned microswimmers}
\label{ss: two aligned microswimmers}

\begin{figure}
    \centering
    \includegraphics[width=0.8\linewidth]{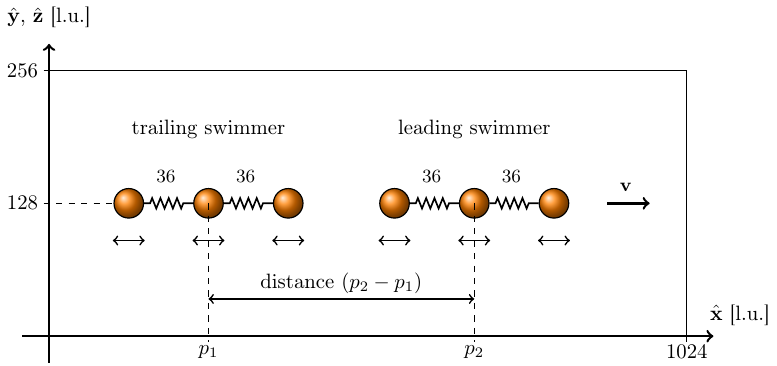}
    \caption{Non-scaled sketch of the in-row configuration for two swimmers. 
    The swimmers move along the x-axis to the right.
    The leading swimmer swims in front of the trailing one. 
    The initial distance between the swimmers is measured as a difference of the center beads within the swimmers $(p_2-p_1)$.}
    \label{fig: s2b sketch}
\end{figure}

\begin{figure}
    \centering
    \includegraphics[width=0.6\linewidth]{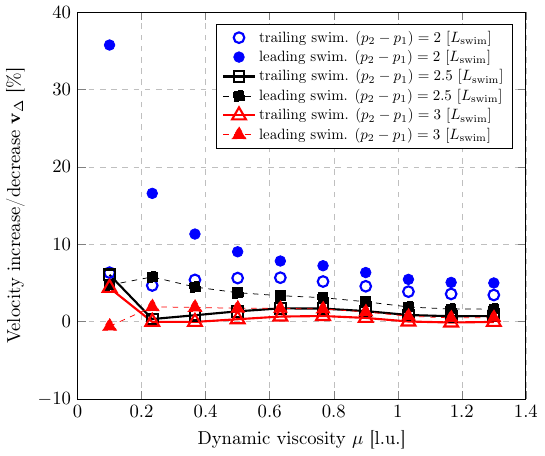}
    \caption{The velocity increase/decrease $\textbf{v}_\Delta = (\textbf{v} - \textbf{v}_\text{ref})/\textbf{v}_\text{ref}$ of two aligned microswimmers for different configurations. 
    The velocities are measured during the 12th swimming cycle (stationary case) with simulation parameters given in Sec.~\ref{s: problem}. 
    The collective motion of two swimmers increases the velocities of both swimmers in certain configurations.}
    \label{fig: s2b vel}
\end{figure}

\begin{figure}
    \centering
    \includegraphics[width=0.9\linewidth]{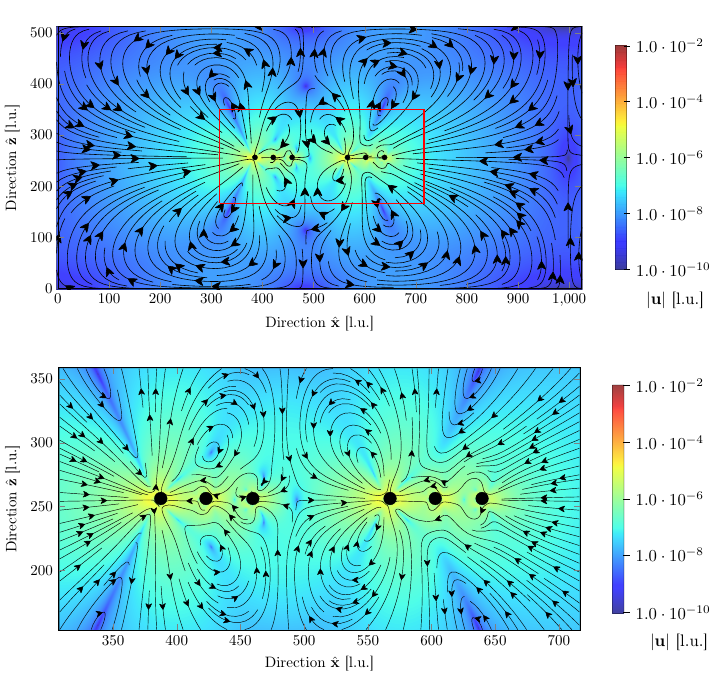}
    \caption{Averaged fluid velocity field for two aligned microswimmers with dynamic viscosity $\mu = 1/6$ and fluid density $\rho_0 =1$. The image above shows the full box and the image below zooms on the microswimmers. The distance between the center beads is $2.5\ L_\text{swim}$ equivalent to $18$ bead diameters.
    The fluid flow (black arrows) surrounding the microswimmers (black beads) is shown whereby it is averaged over one swimming period in the stationary case. 
    The absolute values of the fluid velocity $\textbf{u}$ (black arrows) are depicted by the different colors.}
    \label{fig: s2b vel field}
\end{figure}

Next, we analyze the case of two identical aligned microswimmers, as sketched in Fig. \ref{fig: s2b sketch}. In this case the crucial observable is the relative velocity increase/decrease
\begin{align}
    \displaystyle
    \textbf{v}_{\Delta} = \frac{\textbf{v} - \textbf{v}_\text{ref}}{\textbf{v}_\text{ref}}\,,
    \label{eq: inc}
\end{align}
in comparison to the velocity $\textbf{v}_\text{ref}$ of a single microswimmer. Due to the small velocities of the swimmers, we assume their distance to stay approximately constant over one period. Also the orientation does not change over several periods.\\
Fig.~\ref{fig: s2b vel} shows $\textbf{v}_{\Delta}$ as a function of the dynamic viscosity $\mu$ for different distances among the centers of mass of the swimmers.
As in the single swimmer case, we observe a dependency of the velocity on the fluid's dynamic viscosity.
Interestingly, for a distance of $2\ L_\text{swim}$, both swimmers move faster than a single swimmer, with a velocity increase of up to $36\%$ for the leading swimmer (see Fig.~\ref{fig: s2b vel}).
Since we observe that the leading swimmer moves faster than the trailing one, such configurations are unstable, i.e., the trailing one will be left behind until the distance between their centers of mass reaches $\gtrsim$ 216 [l.u.] = $3\ L_\text{swim}$.
With increasing initial distances between the swimmers, the interaction effect on the swimming velocity of both swimmers decays. 
For instance, the relative velocity increase is below 5 \% for a distance of $3\ L_\text{swim}$ (21.6 bead diameters) in the range of parameters considered (see Fig.~\ref{fig: s2b vel}).
The changes in the interaction behavior as a result of the increasing swimmer distances are negligible in our simulations.
Finally, Fig.~\ref{fig: s2b vel field} shows the fluid velocity field, which is similar to that of a single swimmer shown in Fig.~\ref{fig: s1 vel field}.

\subsection{Two parallel microswimmers}

\begin{figure}
    \centering
    \includegraphics[width=0.8\linewidth]{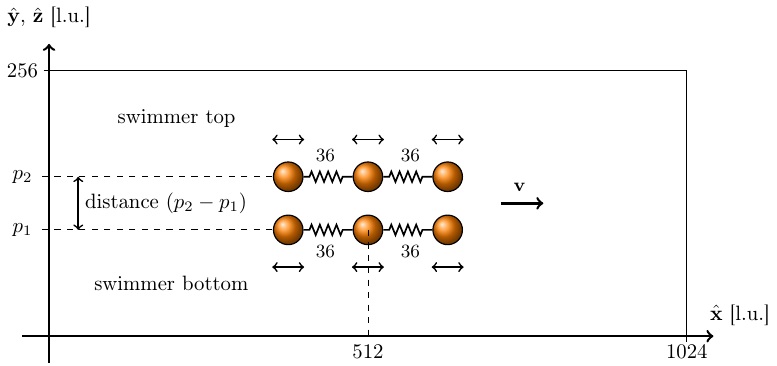}
    \caption{Non-scaled schematics of two microswimmers next to each other. 
    Swimmer 2 is displaced in y-direction by the distance $(p_2 - p_1)$.
    Both swimmers are centered in $\hat{\textbf{y}}$, $\hat{\textbf{z}}$ and swim in $\hat{\textbf{x}}$ direction. 
    The distance between both is measured from the center of mass ($p_1$) of the middle bead of swimmer 1 ($p_1$) to the center of the middle bead of swimmer 2 ($p_2$).}
    \label{fig: s2p sketch}
\end{figure}

\begin{figure}
    \centering
    \includegraphics[width=0.6\linewidth]{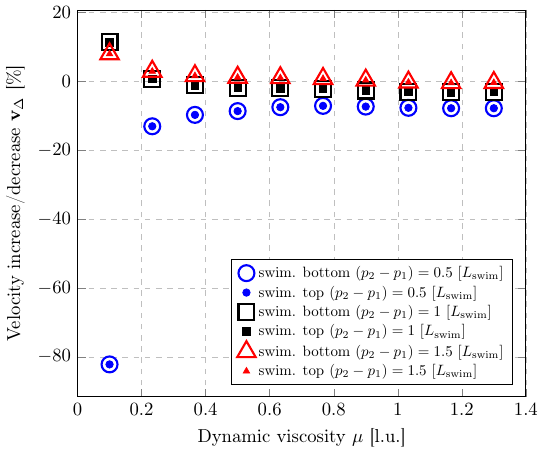}
    \caption{The velocity increase/decrease $\textbf{v}_\Delta = (\textbf{v} - \textbf{v}_\text{ref})/\textbf{v}_\text{ref}$ of two microswimmer next to each other for different configurations.
    The velocities are measured inside the stationary swimming cases, and collective motion implies different velocity increase/decrease effects.
    The transition from decreasing to increasing velocities for larger distances implies an optimal distance for maximal velocity increment.}
    \label{fig: s2p vel}
\end{figure}

\begin{figure}
    \centering
    \includegraphics[width=0.9\linewidth]{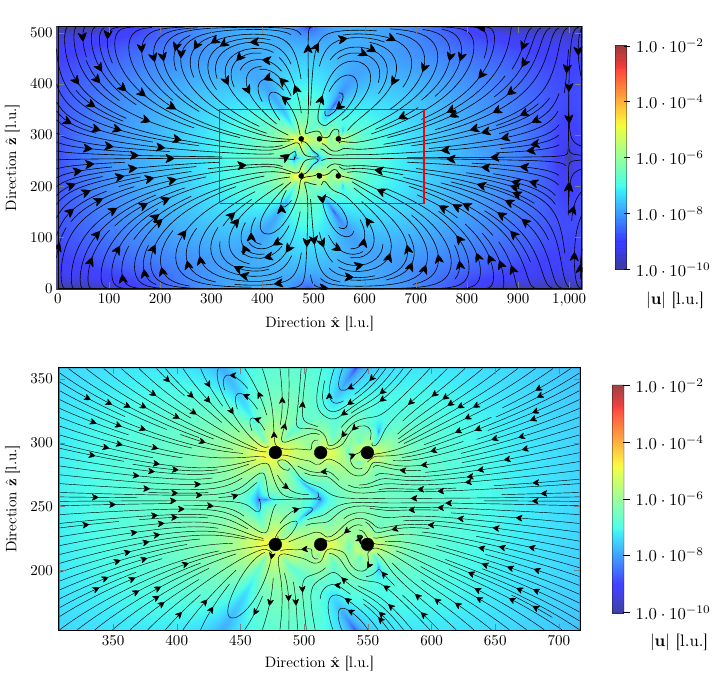}
    \caption{Averaged fluid velocity field for two microswimmers next to each other with dynamic viscosity $\mu = 1/6$ and fluid density $\rho_0 =1$. The image above shows the full box and the picture below zooms on the microswimmers. The distance between the center beads is $1\ L_\text{swim}$ equivalent to $7.2$ bead diameter.
    The fluid flow (black arrows) surrounding the microswimmers (black beads) is shown whereby it is averaged over one swimming period in the stationary case. 
    The absolute values of the fluid velocity $\textbf{u}$ (black arrows) is shown by the different colors.}
    \label{fig: s2p vel field}
\end{figure}

Finally, we discuss the case of two parallel identical swimmers as shown in Fig.~\ref{fig: s2p sketch}.
As expected by symmetry, we find that the two swimmers always have the same velocity for all the distances we investigated among them, i.e., all investigated configurations are stationary on the time scales accessible to lattice Boltzmann simulations. However, theoretical work \cite{PooleyAlexanderYeomans2007, Ziegler2021} has shown that parallel swimmers typically also experience sideways interaction as well as rotation, which will in general break stationarity.\\
Moreover, similarly to the case of aligned swimmers, we observe a dependence of the relative velocity change on the dynamic viscosity of the fluid, as shown in Fig.~\ref{fig: s2p vel}.
Indeed, there is a transition in Fig.~\ref{fig: s2p vel} from decreased velocity to increased velocity between an initial distance of $1\ L_\text{spring}$ and $3\ L_\text{spring}$ which corresponds to 3.6 and 10.8 bead diameters.
This transition implies that swimmers that are close to each other mutually disturb, while swimmers that are further away benefit from the presence of a nearby (yet not too close) companion.
Our data shows that for high viscosity, the velocity increase approaches a constant.
It is reasonable because Fig.~\ref{fig: s1} displays similar behaviour for the motion of a single microswimmer.
When the effect of resonance frequency gets less important, the overall friction reduction is more important for the velocity increase.
Finally, Fig.~\ref{fig: s2p vel field} represents the averaged fluid velocity field.

\section{Discussion}
\label{s: discussion}

\begin{figure}
    \centering
    \includegraphics[width=0.9\linewidth]{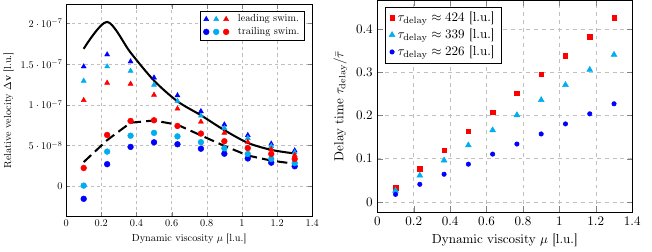}
    \caption{Comparison of Rotne-Prager tensor with delay time, Eq.~\eqref{eq:fit-eq}, and lattice Boltzmann simulations for two microswimmers behind each other with distance two $L_\text{swim}$. The dynamic viscosity $\mu = 1/6$, the fluid density $\rho_0 = 1$ and the length of swimmer is $7.2$ bead diameter. On the left-hand side, the leading (triangle) and trailing (circle) swimmers relative velocity, $v_\Delta$, for diverse proposed delay times are shown. While the black lines are the relative velocities of the leading (smooth) and trailing (dashed) swimmer as obtained from LBM simulations. On the right-hand side, the ratio of the delay time $\tau_\text{delay}$ and the momentum diffusion time $\bar{\tau}= L^2/\nu$ in dependence of the viscosity is shown.}
    \label{fig: delay}
\end{figure}

Comparison of lattice Boltzmann simulations with theoretical approaches based on the Oseen or Rotne-Pragner tensor is generally tricky. 
The theory holds in the Stokes limit, in which the fluid velocity profile adjusts instantaneously to the change of position of the beads. However, in our IBM+LBM+FEM simulations, momentum propagation over a length $L$  occurs on a finite time determined by $\bar{\tau}=L^2/\nu$ with $\nu =\mu/\rho_0$ the kinematic viscosity.
Therefore, the velocity field of one microswimmer needs a finite amount of time to attain its stationary profile and to affect the other beads' motion. In our simulations we have $L\simeq 36 $ [l.u.] = 3.6 [bead diameter] and $\nu \in[0.1,1.2]$ (with $\rho_0=1$) and hence we have $\bar{\tau}\in [1000,12000]$ time steps.
Clearly, for larger values of the viscosity $\bar{\tau}\simeq 1000$, it is reasonable to have a time scale separation between $\bar{\tau}$ and the period of the force $T=20000$. 
In contrast, this is no longer the case for smaller viscosity values, and hence larger deviations between numerical and analytical results are to be expected. 
We have numerically solved the equation of motion for the swimmers using the Rotne-Prager tensor where the delay time has been introduced as an additional parameter. 
Fig.~\ref{fig: delay} shows the dependence of the velocity enhancement of the leading and trailing swimmer as a function of the viscosity. 
In order to discuss the magnitude of the delay time, $\tau_\text{delay}$, we have compared it with $\bar{\tau}=L^2/\nu$.\\
Some preliminary results for the finite relaxation time of the velocity field show an improvement in the analytical/numerical comparison.
Here, the equation of motion in the analytical calculations has been adapted to 
\begin{equation}
    \dot{\mathbf{R}}_i (t) = \frac{1}{6 \pi \mu r_\text{bead}} (\mathbf{f}^d_i (t) + \mathbf{f}^s_i (t)) + \sum_{j \neq i} \hat{T} ( \mathbf{R}_i(t) - \mathbf{R}_j(t - \tau_\text{delay})) \cdot (\mathbf{f}^d_j(t - \tau_\text{delay}) + \mathbf{f}^s_j (t - \tau_\text{delay})),
    \label{eq:fit-eq}
\end{equation}
with $\tau_\text{delay}$ being the time delay. 
Assuming a time delay $\tau_\text{delay}$ independent of the fluid viscosity, we find that good agreement between the LBM and the analytical model can be observed by properly choosing $\tau_\text{delay}$ (left panel of Fig.~\ref{fig: delay}). Also, the obtained delay times are comparable to the time $\bar{\tau}$ that the fluid momentum takes to propagate over half of the swimmer length.\\
The interaction effects of microswimmers can generally be separated in so-called passive effects due to the time-averaged flow field a swimmer is immersed in, and active effects resulting from the interplay of the swimmer's own swimming activity and the time-dependent flow field at the swimmer's position \cite{PooleyAlexanderYeomans2007}. 
While some swimmers, such as the squirmer, only experience passive effects \cite{IshikawaSimmondsPedley2006}, shape-changing swimmers such as the bead-spring swimmer experience both types. 
In particular, bead-spring swimmers will generally alter their swimming stroke as a consequence of the time-dependent flow field they find themselves in, giving rise to active effects \cite{Ziegler2021}. 
These active effects have been reported to be particularly important as small swimmer separations \cite{PooleyAlexanderYeomans2007}, and thus are expected to play an important role in our simulations. It is, in our case, therefore not possible to directly relate the flow field produced by the swimmers (Figs.~\ref{fig: s2b vel field}, ~\ref{fig: s2p vel field}) and the interaction effects experienced by both.

\section{Conclusions}
\label{s: conclusions}

We reported on LBM simulations of one and two bead-spring swimmers. 
As expected, for the single swimmer, we found that its velocity is sensitive to the dynamic viscosity of the fluid, $\mu$, and it shows a non-monotonous dependence on $\mu$ as shown in  Fig.~\ref{fig: s1}.
Our numerical results show a good quantitative agreement with the theoretical predictions based on both the Oseen (Ref. \cite{Pande2015, Pande2017}) or Rotne-Pragner (Ref. \cite{Ziegler2019, Ziegler2021}) tensor as shown in Fig. \ref{fig: s1}.\\
Next, we analyzed the collective motion of two microswimmers in the case in which they are aligned or alongside. 
In the case of aligned swimmers, we found that when the swimmers are close, the speedup due to hydrodynamic interactions of the leading swimmer is more significant than that of the trailing one hence leading to unstable clusters. 
At variance, for the case of alongside swimmers, we found that all the distances among their centers of mass are stationary so that swimmers can proceed together. 
In the case of alongside swimmers, we found that when swimmers are too close, they slow down, whereas they speed up at larger (yet finite) distances. Since when the separation distance diverges we expect no speed up, our results suggest the existence of an optimal length at which the speed up is maximized.\\
Finally, we have commented on the mismatch of the velocity profile as obtained from the LBM simulations and the analytical calculations. Supported by some preliminary results, we argue that this is due to the finite relaxation time of the fluid velocity in the LBM as compared to the Stokes regime assumed in the analytical calculations.\\
While microswimmers such as squirmers or phoretic colloids experience only passive interactions~\cite{IshikawaSimmondsPedley2006}, shape-changing microswimmers as the three-bead swimmer are generally subject to both active and passive interactions~\cite{PooleyAlexanderYeomans2007, Ziegler2021}. 
Active interaction effects have been reported to become particularly important for smaller swimmer separations~\cite{PooleyAlexanderYeomans2007}, as studied in this work. 
Consequently, the interaction behavior presented here will generally differ from that of swimmers of fixed shape, e.g. squirmers. Still, our results might prove useful to understand better the interaction of shape-changing bacteria or algae cells, such as, for instance, Chlamydomonas reinhardtii.

\section*{Acknowledgments}

The DFG Priority Programme SPP 1726 \glqq Microswimmers—From Single Particle Motion to Collective Behaviour\grqq{} (HA 4382/5-1) and SFB 1411 (Project-ID 416229255) financially supported this work. 
We also acknowledge the Jülich Supercomputing Centre (JSC) for the allocation of computing time.

\end{document}